\documentclass[prl,aps]{revtex4}

\usepackage{epsfig}
\usepackage{subfig}
\usepackage{graphicx,color}
\usepackage{amssymb}

\usepackage{amsmath}
\newcommand{\beq}{\begin{equation}}
\newcommand{\eeq}{\end{equation}}
\newcommand{\be}{\begin{eqnarray}}
\newcommand{\ee}{\end{eqnarray}}

\begin{document}

\title{
Discrimination of the light CP-odd scalars
between in the NMSSM and in the SLHM
}

\author{ C. S. Kim }
\email{cskim@yonsei.ac.kr}
\vskip 0.5cm

\affiliation{
Department of Physics and IPAP,
Yonsei University, Seoul 120-749, Korea
}

\author{ Kang Young Lee }
\email{kylee14214@gmail.com}
\vskip 0.5cm

\affiliation{
Division of Quantum Phases \& Devices,
School of Physics, Konkuk University, Seoul 143-701, Korea
}

\author{ Jubin Park }
\email{honolo@phys.nthu.edu.tw}
\vskip 0.5cm

\affiliation{
Department of Physics, National Tsing Hua University, HsinChu 300, Taiwan}

\date{\today}
%\vspace{2cm}
%\maketitle

\begin{abstract}

\noindent The presence of the light CP-odd scalar boson
predicted in the next-to-minimal supersymmetric model
(NMSSM) and the simplest little Higgs model (SLHM)
dramatically changes the phenomenology of the Higgs sector.
We suggest a practical strategy to discriminate the underlying model
of the CP-odd scalar boson
produced in the decay of the standard model-like Higgs boson.
We define the decay rate of ``the non $b$-tagged jet pair"
with which we compute the ratio of decay rates into lepton and jets.
They show much different behaviors between the NMSSM and the SLHM.

\end{abstract}

\pacs{PACS numbers: }

\maketitle

%\end{titlepage}

%\voffset 0cm

%\begin{multicols}{2}
%\narrowtext

%\tightenlines

\section{Introduction}
\label{sec:1}

Discovery of Higgs bosons is the principal goal of the CERN
Large Hadron Collider (LHC), which is essential for understanding
the electroweak symmetry breaking.
In the standard model (SM), there exists one scalar boson $h$
of which mass is constrained by a lower bound of 114 GeV from
the direct search of
$e^- e^+ \to Z h \to Z b \bar{b}$ process
at CERN Large Electron Positron collider (LEP)
\cite{Barate:2003sz}.
At the LHC, the promising production channel is the gluon fusion,
$pp \to gg \to h$. 
The Higgs boson dominantly decays into $WW/ZZ$ when $m_h > 140$ GeV,  
and into $b \bar{b}$ in the low mass region.
Since $h \to b \bar{b}$ channel suffers from huge QCD background,
the favored search channel is $h \to \gamma \gamma$.
The LHC has reported that the accumulated data
of this year exceeds 5 fb$^{-1}$ and more than 10 fb$^{-1}$
is expected in the next year \cite{recentLHC}.
If so, the SM-like Higgs boson might be observed
at the early LHC era with the 7 TeV center-of-mass energy
\cite{higgs}.

However, the ATLAS and the CMS group have examined
the Higgs boson to obtain the null result
in the region $m_h > 140$ GeV with data of 1 fb$^{-1}$ so far.
There remains only the low mass window of 115 GeV$ < m_H < $140 GeV,
which will be narrowed soon with the data of 5 fb$^{-1}$
collected in 2011.
As the LHC will run successfully to accumulate 15 fb$^{-1}$ next year,
it is possible to exclude the whole region of $m_h < 600$ GeV
instead of the observation of the Higgs signal.
If the SM Higgs boson is excluded at the LHC,
it is a clear evidence of the existence of the new physics beyond the SM.
We have to deliberate the implication of the absence
of the SM-like Higgs boson at this stage.
In some scenarios of new physics models beyond the SM,
it is hard to detect the SM-like Higgs boson,
at least at the early stage of the LHC.
One of the examples of this scenario is
the next-to-minimal supersymmetric model (NMSSM)
with an additional singlet superfield.
In the NMSSM,
the parameter space with a very light CP-odd scalar $a$ is allowed
such that the SM-like CP-even Higgs boson can decay
into a pair of light CP-odd scalar bosons with a large branching ratio
\cite{nmssm1}.
Such a light CP-odd scalar sequentially decays into
$b \bar{b}$, $c \bar{c}$, and $\tau^- \tau^+$ pairs depending on its mass.
Then it would be much more difficult to identify the Higgs boson
with four final states and we might miss the Higgs boson signal
at the early LHC.
Moreover in this case,
other channels for Higgs boson decays are suppressed due to
the large branching ratio of $h \to aa$.
On the other hand,
the present Higgs mass bound from the LEP data can be lowered
due to the reduction of the $ZZh$ coupling
and Br$(h \to b \bar{b})$ in this model
\cite{dermisek}.
The simplest little Higgs model (SLHM) with the $\mu$ parameter
is another example of the model with the light CP-odd scalar
\cite{schmaltz}.
The production and decays of the SM-like Higgs boson in the SLHM
are similar to those in the NMSSM,
such as the Higgs boson dominantly decays into
a pair of light CP-odd scalars $\eta$ and the Higgs mass bound weakens
\cite{cheung2006,cheung2007,cheung2008}.

In these scenarios,
the promising channel to find the Higgs boson is
$ h \to aa / \eta \eta \to b \bar{b} b \bar{b}$ via $Wh$
and $Zh$ production at the LHC if $m_{a,\eta} > 2 m_b$,
which is feasible to observe the Higgs boson
through this channel at the LHC with 14 TeV \cite{cheung2007}.
If $m_{a,\eta} < 2 m_b$, it may be produced
in radiative heavy quarkonium decays \cite{domingo}
or in associated production \cite{lee}.
The collider signatures of the CP-odd scalar boson
are similar at the LHC in both models, the NMSSM and the SLHM.
Thus it is not easy to fix the underlying theory
even if we observe a light CP-odd scalar boson at the LHC.
Therefore it is very important to clarify the underlying structure
of the light CP-odd scalar boson.
In this Letter,
we present a strategy to discriminate the underlying
model of the CP-odd scalar boson
assuming that the CP-odd scalar boson has been already discovered
through $h \to aa / \eta \eta$ decays
and its mass is measured.
Since the producton cross sections and the decay rates depend upon
many model parameters as well as the final state masses,
we cannot fix the model by the measured cross sections
and branching ratios.
If we accumulate enough number of CP-odd scalars
to estimate the ratio of decay rates,
however, the most parameter dependences are canceled
and the features of the underlying models are revealed.
Especially in the SLHM,
the Yukawa couplings are commonly expressed by the new scale
$f$ and $\tan \beta$ and the ratios of decay rates are
determined by the $\eta$ mass and the final states masses only.

We consider the ratio of the decay rates into tau lepton pair
to those into quark pairs.
Since it is impossible to identify the $c$-jet
and only partly possible to tag the $b$-jet,
we define a new observable $\Gamma(j' j')$,
the decay width into ``non $b$-tagged jet pair",
by subtracting tagged $b$-jet from the total jet decay rates.
We show that it is possible to discriminate the underlying models
of the CP-odd scalars with the ratio of
the decay rates into tau lepton pair and into non $b$-tagged jet pair
with an allowed value of the $b$-tagging efficiency.
If we will not observe the SM-like Higgs boson at the early stage
of the LHC, the light CP-odd scalar scenario should be
concerned seriously.
Then it will be very important to find out the underlying model
of the CP-odd scalar.

\section{Two scenarios for the light CP-odd scalar from the NMSSM and the SLHM}
\label{sec:2}

In the NMSSM, the Higgs sector is described by the superpotential
\cite{nmssm},
\be
W = \hat{Q} \hat{H}_u h_u \hat{U}^c
+ \hat{H}_d \hat{Q} h_d \hat{D}^c + \hat{H}_d \hat{L} h_e \hat{E}^c
+ \lambda \hat{S} (\hat{H}_u \hat{H}_d) + \frac{1}{3} \kappa \hat{S}^3,
\ee
where $\hat{S}$ is a singlet chiral superfield.
The associated soft trilinear couplings are given by
\be
V = \lambda A_\lambda S H_u H_d + \frac{1}{3} \kappa A_\kappa S^3 + H.c.~,
\ee
The effective $\mu$ term is generated by the vacuum expectation value (VEV)
of the singlet scalar $s \equiv \langle \hat{S} \rangle$,
yielding $\mu = \lambda s$.
With an extra complex singlet scalar field,
the Higgs sector of the NMSSM consists of
three CP-even Higgs bosons, two CP-odd Higgs bosons,
and a pair of charged Higgs boson
and is described by six parameters
$\lambda$, $\kappa$, $A_\lambda$, $A_\kappa$,
$\tan \beta$, and $\mu_{\rm eff}$
where $\tan \beta = \langle H_u \rangle/\langle H_d \rangle$.
The relevant Lagrangian for CP-odd scalars is given by
\be
\mathcal{L}=i\frac{g}{2m_{W}} p_{\,i1}
      \Big[ m_{d}\tan\beta~ \bar{d}\gamma_{5}d
           + m_{u}\cot\beta~ \bar{u}\gamma_{5}u\Big] P_{i} ~,
\ee
where $\tan \beta = v_1/v_2$ and
$p_{\,i1}$ are mixing matrix elements
between pseudoscalar components $P_{1}$ and $P_{2}$
dropping the Goldstone mode.
We introduce an angle $\gamma$ to represent the mixing such that
\be
\left(
  \begin{array}{c}
    P_{1} \\
    P_{2} \\
  \end{array}
\right) =
\left(
  \begin{array}{cc}
    \cos\gamma & -\sin\gamma \\
    \sin\gamma & \cos\gamma \\
  \end{array}
\right)
\left(
  \begin{array}{c}
    a \\
    A \\
  \end{array}
\right)~,
\ee
where $a$ is the physical state of the light CP-odd scalar
and $A$ that of the heavy one.
We concentrate only on $a$ in this Letter.

The light CP-odd Higgs boson arises in the SLHM with the $\mu$ term,
where the global symmetry is $[SU(3) \times U(1)_{X}]^2$
wiht the gauge symmetry $SU(3) \times U(1)_{X}$
as its diagonal subgroup.
The symmetries are broken to the SM gauge symmetry
by the VEV of the non-linear SU(3) triplet scalar fields,
$\langle \mathbf{\Phi_{1,2}} \rangle =(0,0,f_{1,2})^T$.
We assume that $f_{1,2}$ are of order TeV.
The remnant degrees of freedom in the Goldstone boson sector
are the $SU(2)_L$ doublet $H$ and a CP-odd scalar boson $\eta$.
The Higgs potential is radiatively generated
via fermion and gauge boson loops
and so is the SM-like Higgs boson mass.
However, the $\eta$ remains massless because
it appears in the phase factor of $\mathbf{\Phi_{1,2}}$.
The massless $\eta$ has a trouble with constraints
from the rare $K$, $B$ decays, radiative $\Upsilon$ decays, and cosmology.
Thus we introduce
$ -\mu^{2}(\mathbf{\Phi_{1}^{\dagger}\Phi_{2}}+ H.c.)$ term
to generate the mass term of CP-odd scalar,
even though it breaks the global symmetry slightly.
The fermion doublets of the SM are promoted to the $SU(3)$ triplets
in this model.
In addition, heavy fermions are required
in order to cancel quadratic divergences of SM top quark
and remove the anomaly of the gauge group.
%Note that there are two kinds of embedding of
%these heavy fermions in the literature.
%One is the "universal" embedding and
%the other one is the "anomaly-free" embedding.
%In this paper we only focus on the anomaly-free embedding.
The relevant Lagrangian with Yukawa interactions is
\be
\mathcal{L}=-i\sum_{f}\frac{m_{f}}{v}~y_{f}^\eta~ \eta~ \bar{f}\gamma_{5}f
      + \frac{m_{t}}{v} (i \eta \bar{T}P_{R}t+ h.c)~,
\ee
where the Yukawa couplings are
\be
&&y^{\eta}_{l}=y^{\eta}_{d, s, b}=-y^{\eta}_{u, c, t}
                    =\frac{\sqrt{2}v}{f}\cot 2\beta,
\nonumber \\
&&y^{\eta}_{Q}=-\frac{v}{f}[\cos 2\beta + \cos 2\theta_{Q}]\csc 2\beta~,
\ee
with leptons $l=e, \mu, \tau$, heavy quarks $Q=D,S,T$.
In this model, the new physics scale $f^2 = f_1^2+f_2^2$
and $\tan \beta = f_2/f_1$,
and the mixing angles $\theta_{T,S,D}$
between the heavy quarks $T,S,D$ and SM quarks $t,s,d$
are defined by
\be
\cos 2\theta_{F}=\sqrt{1-\frac{2f^{2}}{v^{2}}
            \frac{m_{q}^{2}}{m_{Q}^{2}}\sin^{2}2\beta}~.
\ee

\section{Discriminatory Signatures of Two light CP-odd scalars}
\label{sec:3}

In our scenario, we assume that the light CP-odd scalar is produced
by the SM-like Higgs boson decay, $h \to aa / \eta \eta$.
Then the CP-odd scalar will decay into fermion pairs and gauge boson pairs,
and the decay channels depend on its mass.
The decays of $\eta$ in the SLHM are given in Ref. \cite{cheung2008}
and those of $a$ in the NMSSM given in Ref. \cite{nmssm}.
Since the Yukawa couplings are proportional to the fermion masses,
the dominant decay modes are
$b \bar{b}$, $c \bar{c}$, and $\tau^- \tau^+$
when $m_{a,\eta} > 2 m_b$.
However, it is not likely to identify the $c$-quark jets at the LHC,
we consider the decays into two jets instead.
We define the ``non $b$-tagged jet pair'', $ j' j'$,
which is two jet event not tagged as $b$-quark jets.
If we let the $b$-tagging efficiency as $\epsilon_b$,
the decay rate of the non $b$-tagged two jet event is
obtained by subtracting $b$-tagged events from total two jet events, $i.e.$
\be
\Gamma(j' j') = \Gamma(b \bar{b}) (1- \epsilon_b) + \Gamma(c \bar{c})
              = \Gamma(jj) - \epsilon_b \Gamma(b \bar{b}),
\ee
where $\Gamma(jj)$ is the total decay width of the CP-odd scalar
into two jets.

In the NMSSM, the decay width of $a$ to down type $f\bar{f}$ is given by,
\be
\Gamma(a \rightarrow f\bar{f})=\frac{N_{C}}{8\pi}
             \Big(\frac{m_{f}}{v}\Big)^{2} C_\beta^2 \, \cos^{2}\gamma ~m_{a}
             \Bigg(1-\frac{4m_{f}^{2}}{m_{a}^{2}}\Bigg)^{\frac{1}{2}}~,
\ee
where $C_\beta = \tan \beta$ for down-type quarks and leptons, and
$C_\beta = \cot \beta$ for up-type quarks.
%The decay rate of $a$ to $\gamma\gamma$ is given by
%\cite{chang},
%\be
%\Gamma(a \rightarrow \gamma\gamma)=\frac{9\alpha^{2}}{1024\pi^{3}m_{f}^{2}}
%       \Bigg(\frac{m_{f}}{v}\Bigg)^{2} \cot^{2}\beta\,
%               \cos^{2}\gamma\, b_{i}^{2} m_{a}^{3}~,
%\ee
%where $b_{i}$ is the contribution of the vectorlike fermion
%to the beta function for given gauge group.

\begin{figure}[t!]
%\centering
\begin{center}
\subfloat{
\includegraphics[width=8cm]{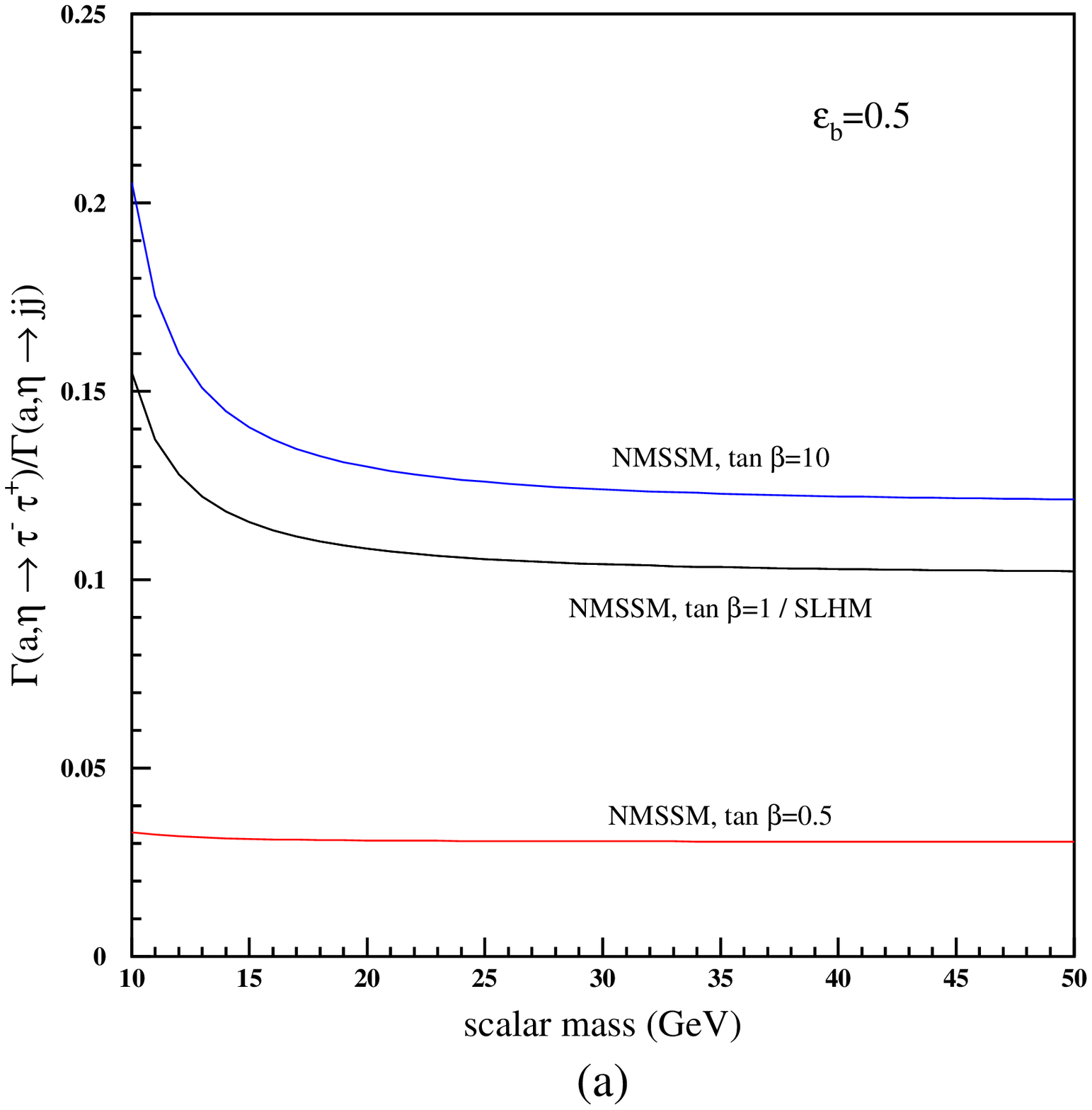}
}
\subfloat{
\includegraphics[width=8cm]{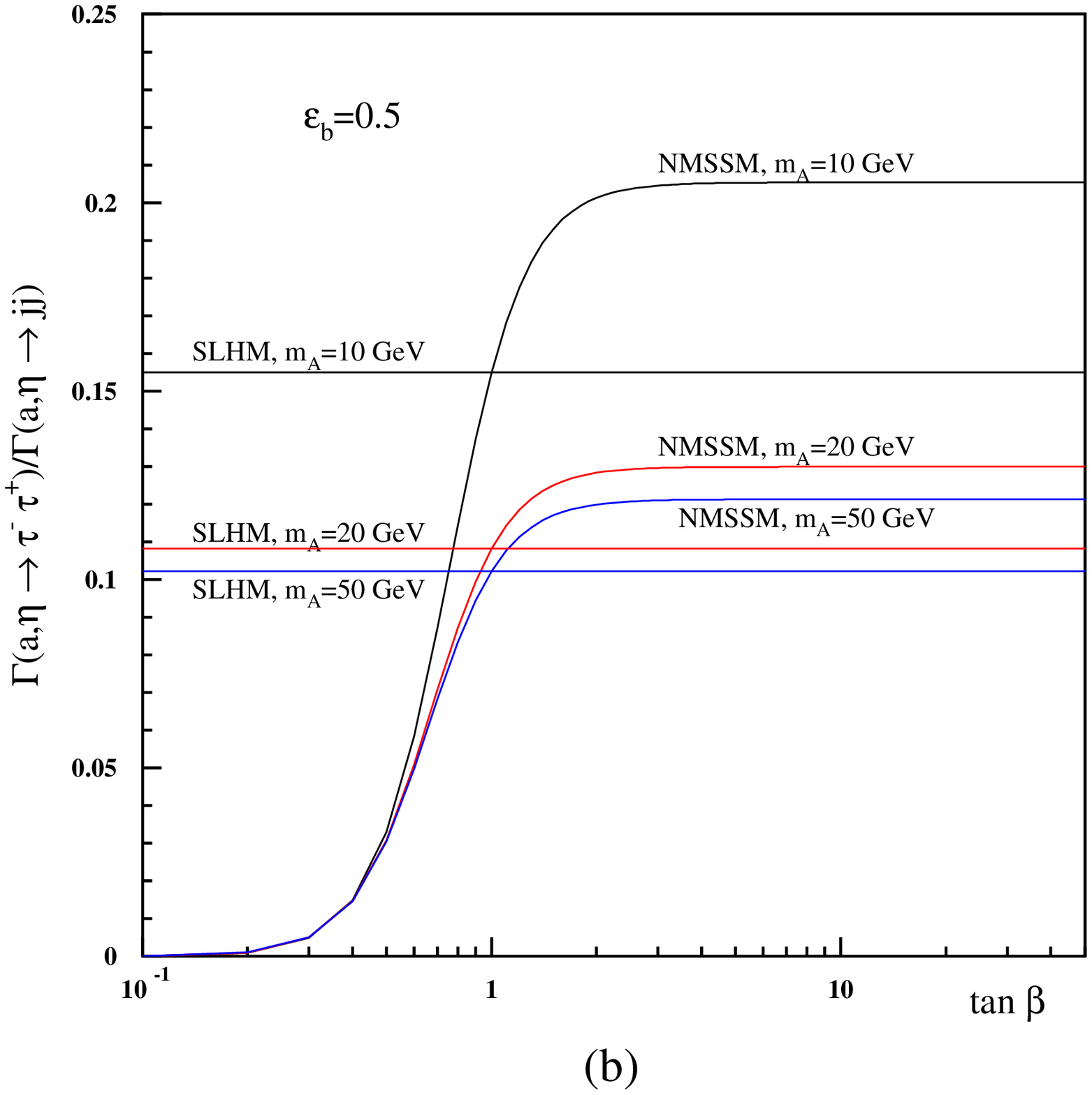}
}
\end{center}
\caption{\label{fig1}\small
The ratios of decay rates into $\tau^- \tau^+$ and
into the non $b$-tagged two jet as functions of 
(a)the CP-odd scalar mass and (b) $\tan \beta$.
}
\end{figure}

If we consider the ratio of $b \bar{b}$ and $\tau^- \tau^+$,
$\tan \beta$ dependence is canceled and
the ratio is determined by only the final state masses and $m_a$,
as is the case of the SLHM shown below.
Thus we cannot discriminate two models.
Instead we consider the ratio of $\tau^- \tau^+$ and
non $b$-tagged two jet events,
which reveals $ \tan \beta$ dependence, as given by
\be
\frac{\Gamma(a \to \tau^- \tau^+)}{\Gamma(a \to j' j')}=
    \frac{ m_{\tau}^{2} f(m_\tau)}
    {3 m_{b}^{2} f(m_b) (1-\epsilon_b)+3 m_{c}^{2} \cot^4 \beta f(m_c)} ,
\ee
where $f(m) = \sqrt{1-4m^2/m_a^2}$.
We note that the additional parameter $\cos \gamma$
does not appear in the ratio of decay rates.

In the SLHM, the decay rates of $\eta$ into fermion pairs are given by
\be
\Gamma(\eta \rightarrow f\bar{f})=\frac{N_{C}}{8\pi}\Big(\frac{m_{f}}{v}
              \Big)^{2} y^{\eta\,2}_{f} m_{\eta}
           \Bigg(1-\frac{4m_{f}^{2}}{m_{\eta}^{2}}\Bigg)^{\frac{1}{2}}~,
\ee
We now consider the ratios of their decay widths into fermions;
\be
\frac{\Gamma(\eta \rightarrow f\bar{f})}{\Gamma(\eta \rightarrow f'\bar{f'})}=
    \frac{N_{C}~ m_{f}^{2} \Big(1-4m_{f}^{2}/m_{\eta}^{2}\Big)^{\frac{1}{2}}}
        {N_{C}~ m_{f'}^{2} \Big(1-4m_{f'}^{2}/m_{\eta}^{2}\Big)^{\frac{1}{2}}}~.
\ee
Note that the ratio is determined by only the final state fermion masses
and scalar mass and does not depend on any model parameters.
Thus the ratio of the SLHM is identical to that of the NMSSM 
with $\tan \beta=1$.

Figure 1 (a) depicts the ratios 
$\Gamma(a \to \tau^- \tau^+)/\Gamma(a \to j' j')$ 
as functions of $\tan \beta$
in the NMSSM and the SLHM.
Note that the ratio of the SLHM falls on 
that of $\tan \beta=1$ case in the NMSSM.

\begin{figure}[t!]
\centering
\subfloat{
\includegraphics[width=6cm]{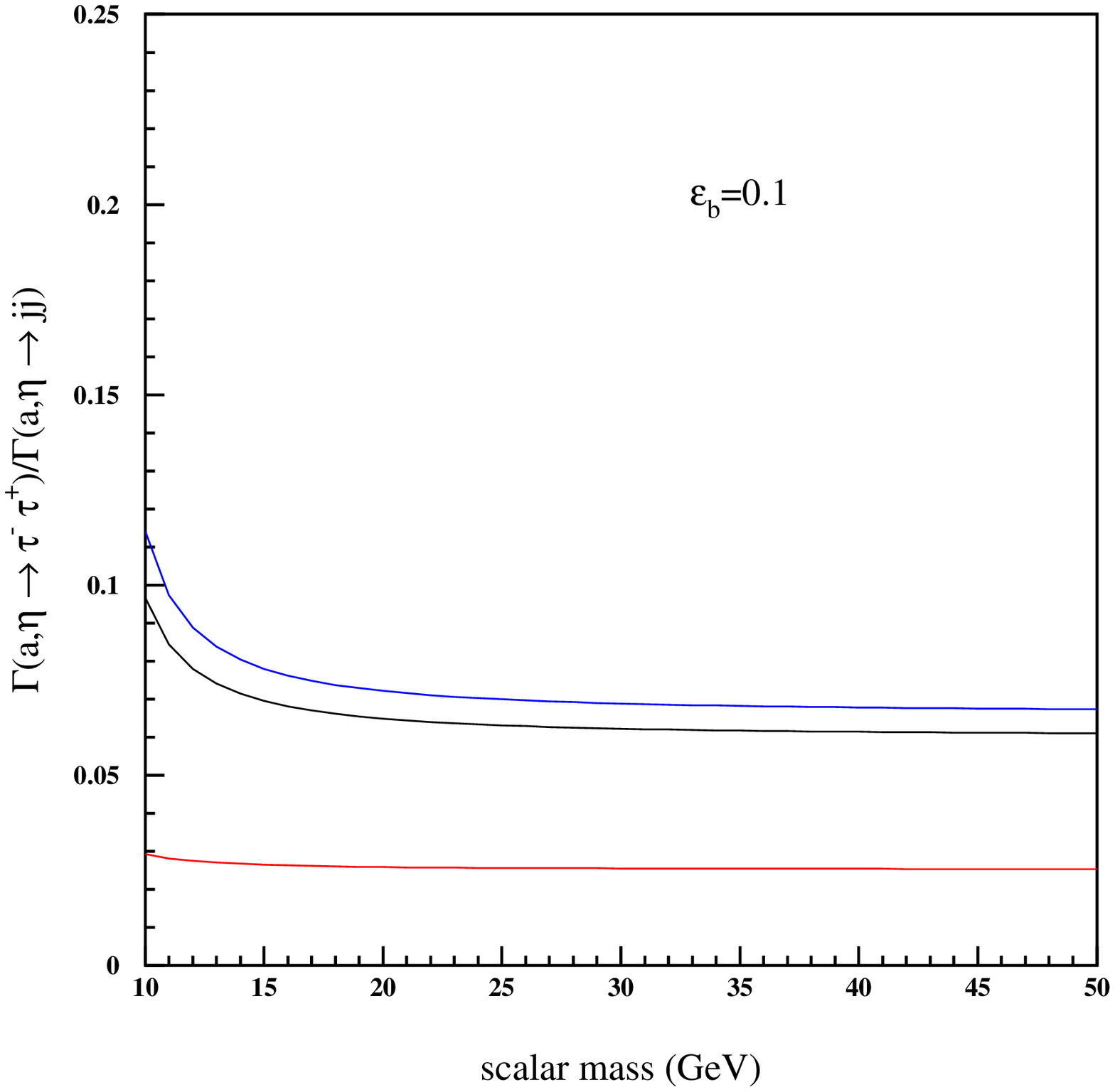}
}
\subfloat{
\includegraphics[width=6cm]{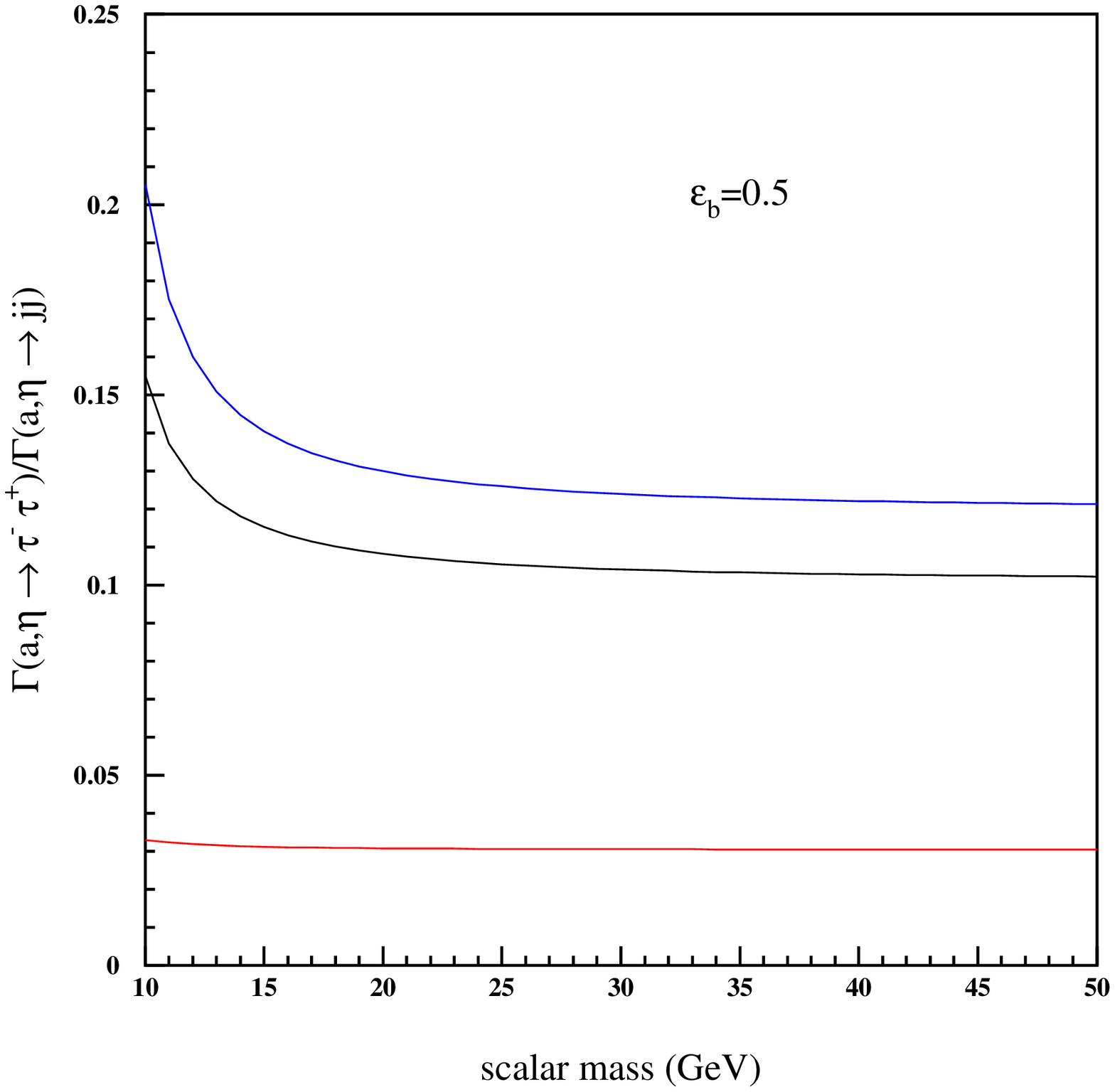}
}
\subfloat{
\includegraphics[width=6cm]{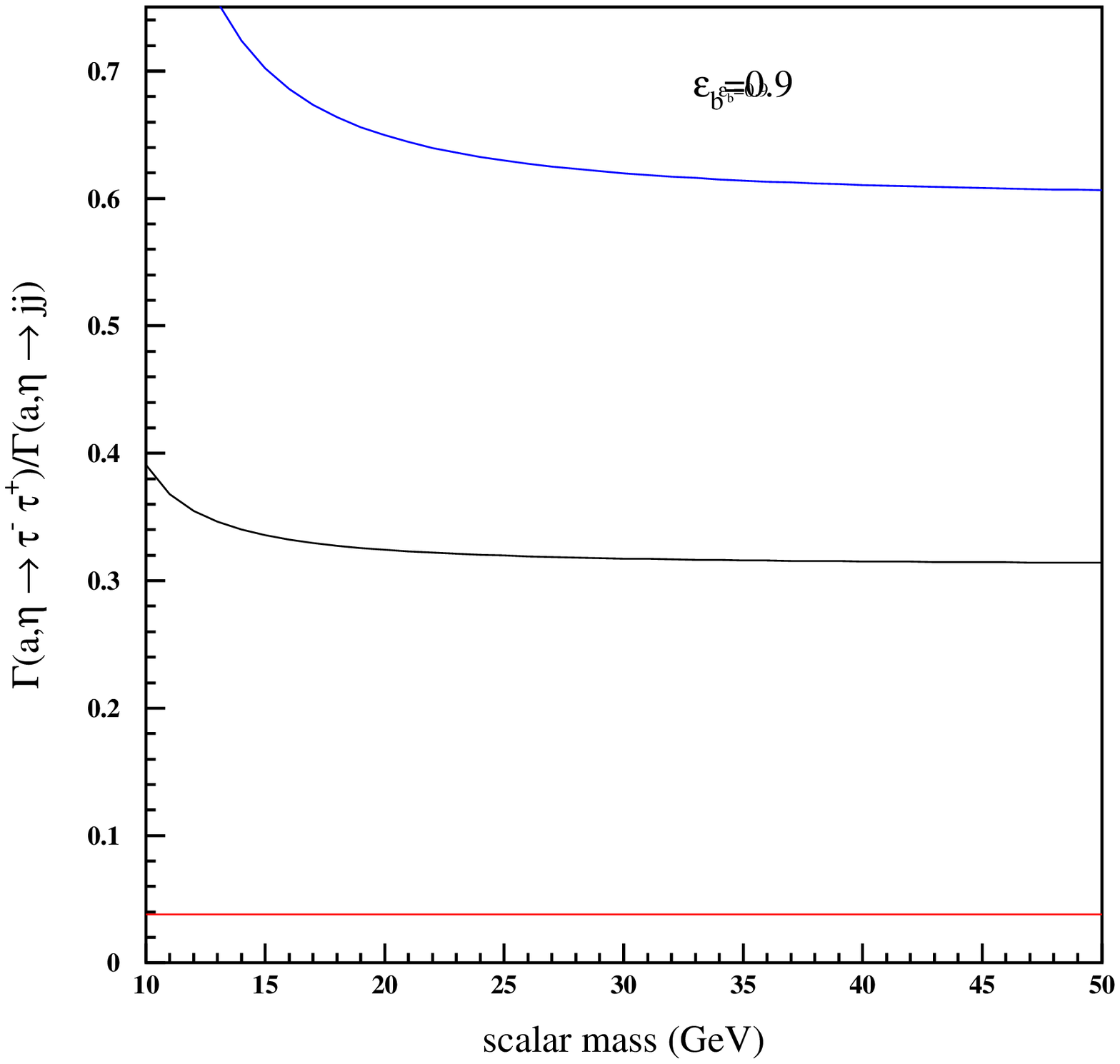}
}
\caption{\label{fig2}\small
The ratios of decay rates into $\tau^- \tau^+$ and
into the non $b$-tagged two jet as functions of 
the CP-odd scalar mass
with respect to
the $b$-tagging efficiency, $\epsilon_b=0.1, 0.5, 0.9$.
The upper curve (blue) denote the NMSSM with $\tan \beta=10$, 
the middle curves (black) the NMSSM with $\tan \beta=1$ and the SLHM, 
and the lower curves (red) the NMSSM with $\tan \beta=0.5$ 
in each plot.
}
\end{figure}

Figure 1 (b) depicts the $\tan \beta$ dependence of the ratio.
If $\tan \beta > 1$, the $c \bar{c}$ contribution becomes
less important and the ratio is saturated.
The predictions for the SLHM do not depend on $\tan \beta$.
Instead the SLHM plot crosses the NMSSM curves at $\tan \beta =1$
for all $m_\eta$.
Please note that the definitions of $\tan \beta$ in the NMSSM and the SLHM
are different from each other.

We assume that 
we can observe an CP-odd scalar boson from the SM-like Higgs decay,
and collect data enough for determination of branching ratios.
If the origin of the CP-odd scalar is the SLHM,
the ratio should be fixed, $e.g.$ $\sim 0.11$ when $m_{\eta} > 20$ GeV.
If we measured the ratio much different from that value,
we can exclude the SLHM, and the NMSSM is a strong candidate.
If $m_{a,\eta} \to 2 m_b$, the kinematic factor for
$b \bar{b}$ channel vanishes and the ratio is close to 1/3
due to the color factor.

In Fig. 1, we set $\epsilon_b=0.5$. 
The $b$-tagging is achieved by several algorithms at the LHC,
e.g. track counting, simple secondary vertex, and their variants.
If we cannot tag $b$-jets, $i.e.$ $\epsilon_b=0$, the $b$-quark contribution
dominates in the denominator 
and the $\tan \beta$ dependence of the ratio of Eq. (10) 
is very weak when $\tan \beta >1$.
Then it is hard to discriminate the NMSSM from the SLHM in this region.
If $\epsilon_b=1$, we can tag all $b$-jets, and we obtain the ratio
$\Gamma(\tau \tau)/\Gamma(c \bar{c})$ to discriminate two models clearly,
which it is an ideal case.
We show the effects of the $b$-tagging efficiency
on the $\tan \beta$ dependence of the ratio in Fig. 2.
We can see that better the $b$-tagging, easier to discriminate the curves.
The recent estimation of the CMS group 
tells us that the efficiency can reach 0.562 \cite{cms-b}.
We find that it is conservatively possible to discriminate two models
with the allowable valus of the$b$-tagging efficiency.

The $a/\eta \to \gamma \gamma$ channel might be useful 
to find the signal due to relatively low background
\cite{chang}.
However, the decay rates are too small to be measured 
$\sim 10^{-4}$,
and even worse to involve many model parameters.
Thus we do not consider this channel in this work.

%%%%%%%%%%%%%%%%%%%%%%%%%%%%%%%%%%%%%%%%%%%%%%%%%%%%%%%%%%%%%%%%%%%%%%%%%%%%

\section{Concluding remarks}

The light CP-odd scalar boson with the dominant
$h \to aa / \eta \eta$ decay provides a new phenomenology
of the Higgs sector.
We have to find the SM-like Higgs boson through identifying $a$ or $\eta$
owing to the large Br$(h \to aa / \eta \eta)$.
In this Letter, assuming that we have observed a CP-odd scalar boson,
we present a strategy to determine it to be
$a$ or $\eta$, the CP-odd scalar in the NMSSM or in the SLHM.
The signal cross section and decay rates depend upon
many undetermined parameters.
Since the Yukawa couplings in the SLHM involve common dependence
on model parameters, the ratios of the decay rates are
expressed by final state masses and $m_a$.
However, in the NMSSM, the Yukawa couplings for the up-type quarks
involve $\cot \beta$, and those of down-type quarks and charged leptons
involve $\tan \beta$ due to the supersymmetry.
Therefore, the ratio of decay rates in the NMSSM can
strongly depend on $\tan \beta$ and show much difference from
that in the SLHM with fixed particle masses.
We define the decay width of non $b$-tagged two jets events,
where the $\tan \beta$ dependence remains in the practical reason.
In conclusion,
if we measure the ratio of decay rates into $\tau$ pair and
non $b$-tagged two jets,
we can easily discriminate two models in the case of
$\tan \beta$ far from 1.

\vskip 1cm

\acknowledgments
C.S.K. was supported by the National Research Foundation
of Korea (NRF) grant funded by the Korean Ministry of
Education, Science and Technology (MEST)
(No. 2011-0027275), (No. 2011-0017430) and (No. 2011-0020333).
K.Y.L. was supported
by WCU program through the KOSEF funded by the MEST (R31-2008-000-10057-0)
and the Basic Science Research Program through the NRF
funded by MEST (2010-0010916).
J. P. was supported by the Taiwan NSC under Grant 
No. 100-2811-M-007-030 and 099-2811-M-007-077.

%%%%%%%%%%%%%%%%%% References
%%%%%%%%%%%%%%%%%%%%%%%%%%%%%%%%%%%%%%%%%%%%%%%%%%%%%%%
\def\PRD #1 #2 #3 {Phys. Rev. D {\bf#1},\ #2 (#3)}
\def\PRL #1 #2 #3 {Phys. Rev. Lett. {\bf#1},\ #2 (#3)}
\def\PLB #1 #2 #3 {Phys. Lett. B {\bf#1},\ #2 (#3)}
\def\NPB #1 #2 #3 {Nucl. Phys. B {\bf #1},\ #2 (#3)}
\def\ZPC #1 #2 #3 {Z. Phys. C {\bf#1},\ #2 (#3)}
\def\EPJ #1 #2 #3 {Euro. Phys. J. C {\bf#1},\ #2 (#3)}
\def\JHEP #1 #2 #3 {JHEP {\bf#1},\ #2 (#3)}
\def\JKPS #1 #2 #3 {J. Korean Phys. Soc. {\bf#1},\ #2 (#3)}
\def\IJMP #1 #2 #3 {Int. J. Mod. Phys. A {\bf#1},\ #2 (#3)}
\def\MPL #1 #2 #3 {Mod. Phys. Lett. A {\bf#1},\ #2 (#3)}
\def\PTP #1 #2 #3 {Prog. Theor. Phys. {\bf#1},\ #2 (#3)}
\def\PR #1 #2 #3 {Phys. Rep. {\bf#1},\ #2 (#3)}
\def\RMP #1 #2 #3 {Rev. Mod. Phys. {\bf#1},\ #2 (#3)}
\def\PRold #1 #2 #3 {Phys. Rev. {\bf#1},\ #2 (#3)}
\def\IBID #1 #2 #3 {{\it ibid.} {\bf#1},\ #2 (#3)}

\end{document}